# Density of electron states in a rectangular lattice under uniaxial stress

Ryszard Piasecki[*]

Institute of Chemistry, University of Opole, Oleska 48, 45-052 Opole, Poland

**Abstract**

The closed analytical expression for the electron density of states function in a rectangular lattice is derived in an elementary way in terms of complete elliptic integrals of the first kind. The lattice can be treated as a deformed square lattice under uniform uniaxial stress (or strain). In contrast to hydrostatic case the uniaxial pressure, say along axis *y*, modifies a length of the *y*-bonds while the *x*-bonds remain intact. It also alters the corresponding tight-binding transfer integral $\gamma_2$ between two *y*-nearest-neighbours leaving unchanged the $\gamma_1$ for *x*-nn interactions. Due to stress-induced lowering symmetry of this simple model one can get an insight into the decoupling of its density of states on dependence of the lattice deformation or transfer integrals anisotropy.

PACS 71.20.-b Electron density of states

## 1. Introduction

Numerous physical properties of crystalline materials under the uniaxial pressure (stress or strain) depend, either directly or indirectly, on the changes of the density of electron states (DOS) in a deformed crystal lattice. One of the recent examples is $CaLi_2$. Its two hexagonal and cubic polymorphs each undergo a structural bifurcation on densification to structures compressed or elongated in a one lattice direction [1]. The suggestion of its potential bulk superconductivity under pressure has been recently confirmed [2].

From theoretical viewpoint, some low-dimensional and quasi-low-dimensional systems are particularly interesting due easily accessible analytic results. Among them are systems enabling for calculations of exact DOS as planar unilaterally modulated superlattices in the presence and the absence of the magnetic field [3,4] or zigzag and armchair single-wall carbon nanotubes [5] with hexagonal Brillouin zone, just to mention a few. The universal expression for DOS in the vicinity of the Fermi level for undeformed carbon nanotubes [6] can be modified to incorporate stretching, compression, torsion and bending [7,8]. The usual one-parameter nearest-neighbour tight-binding model [9,10] was adopted by above workers or extended to more parameters as for graphene nanostrips [11,12].

In general, one can state that the knowledge of analytical DOS formulas is no doubt of some importance. Here, we show a simple way for the exact evaluation of uniaxial pressure-induced changes of this quantity in a deformed square lattice.

## 2. Model and results

Let us consider the energy dispersion relation given by the tight-binding model with the overlap between two nearest-neighbours for atomic *s*-orbitals in a rectangular lattice

$$E(k_x, k_y) = E_0 - 2\gamma_1 \cos(k_x a) - 2\gamma_2 \cos(k_y b) . \quad (1)$$

Note that also $p_z$-orbitals put on the rectangle lattice with the direction perpendicular to the lattice taken as *z* will give a band structure similar to that of *s*-orbital since the topology of the interaction of these orbitals is similar [13].

A reference energy is denoted as $E_0$, and the allowed components $k_x$ and $k_y$ range within $0 \leq |k_x| \leq \pi/a$ and $0 \leq |k_y| \leq \pi/b$. A square lattice is deformed by a uniform uniaxial stress along axis *y* that induces a decrease of the lattice constant *b* and increase of the corresponding nearest-neighbour transfer integral $\gamma_2$. Thus we have the following inequalities for the lattice constants, $a \geq b$, and for the nearest-neighbour hopping parameters, $0 < \gamma_1 \leq \gamma_2$. The used later on a related parameter $0 < \Delta \equiv \gamma_1/\gamma_2 \leq 1$ is indirectly linked with the applied stress. So the difference $1-\Delta$ can be treated as a measure of the lattice deformation (or an anisotropy of the nearest neighbour interactions). At last for a dimensionless energy variable $\varepsilon \equiv (E-E_0)/2\gamma_2$ we obtain a suitable range $-1-\Delta < \varepsilon < 1+\Delta$.

The density of the electron states $g(\varepsilon, \Delta)$ in an energy band for a deformed planar lattice can be expressed as follows

[*] Fax: +48 77 4410740.
 *E-mail address*: piaser@uni.opole.pl (R. Piasecki).

$$g(\varepsilon;\Delta) = \frac{ab}{4\pi^2} \iint_B \delta[E - E(k_x,k_y)] dk_x dk_y$$

$$= \frac{ab}{2\gamma_2\pi^2} \int_0^{\pi/b}\int_0^{\pi/a} \delta[\varepsilon + \Delta\cos(k_x a) + \cos(k_y b)] dk_x dk_y \quad (2)$$

where due to the symmetry of Brillouine zone ($B$) the integrations extend over $0 \leq k_x \leq \pi/a$ and $0 \leq k_y \leq \pi/b$. Substituting $x \equiv \cos(k_x a)$ and $y \equiv \cos(k_y b)$ we obtain

$$g(\varepsilon;\Delta) = \frac{1}{2\gamma_2\pi^2} \int_{-1}^{1}\int_{-1}^{1} \frac{\delta(\varepsilon + x\Delta + y) dx dy}{\sqrt{(1-x^2)(1-y^2)}}$$

$$= \frac{1}{2\gamma_2\pi^2} \begin{cases} \int_I f(x)dx & \text{if } 1-\Delta < \varepsilon < 1+\Delta \\ \int_{II} f(x)dx & \text{if } 0 < \varepsilon < 1-\Delta \\ \int_{III} f(x)dx & \text{if } -1+\Delta < \varepsilon < 0 \\ \int_{IV} f(x)dx & \text{if } -1-\Delta < \varepsilon < -1+\Delta \end{cases} \quad (3)$$

where

$$f(x) = \frac{1}{\sqrt{(1-x^2)[1-(\varepsilon+x\Delta)^2]}}, \quad (4)$$

and the consecutive intervals of integration are equal to $I \equiv \{-1 \leq x \leq (1-\varepsilon)/\Delta\}$, $II \equiv III \equiv \{-1 \leq x \leq 1\}$, and $IV \equiv \{(-1-\varepsilon)/\Delta \leq x \leq 1\}$. For a fixed $\Delta$, also the corresponding parameter spaces $I$, $II$, $III$ and $IV$ are illustrated in Fig. 1.

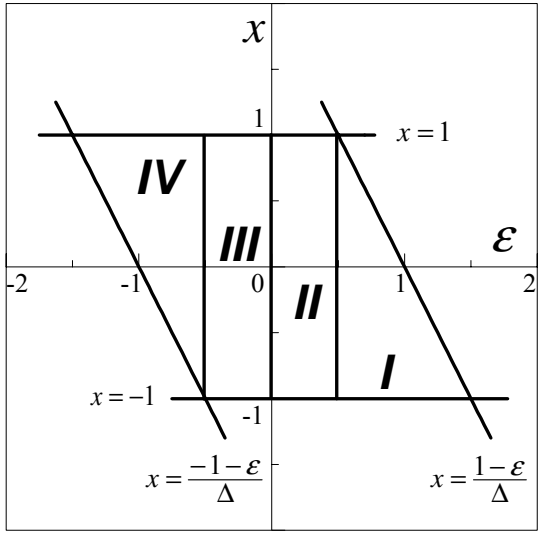

**Figure 1** The ($\varepsilon$, $x$) areas with a fixed parameter $\Delta$, which relate to the four integrals defined in Eq. (3).

In the following, we shall denote the four integrals in Eq. (3) as $g_I(\varepsilon>0; \Delta)$, $g_{II}(\varepsilon>0; \Delta)$, $g_{III}(\varepsilon<0; \Delta)$ and $g_{IV}(\varepsilon<0; \Delta)$, respectively. Note, that for a given $\Delta$ one finds $f(-x; -\varepsilon) = f(x; \varepsilon)$. For this reason it is easy to see that $g_I(\varepsilon>0; \Delta) = g_{IV}(\varepsilon<0; \Delta)$ and $g_{II}(\varepsilon>0; \Delta) = g_{III}(\varepsilon<0; \Delta)$. Thus, without loss of generality one can consider only the two integrals, $g_I(\varepsilon>0; \Delta)$ and $g_{II}(\varepsilon>0; \Delta)$. To integrate the function $f(x)$ it is convenient to rewrite it as

$$f(x) = \frac{1}{\Delta\sqrt{\prod_{i=1}^{4}(x-x_i)}} \equiv \frac{1}{\Delta\sqrt{X}} \quad (5)$$

where $x_1 = 1$, $x_2 = -1$, $x_3 = (1-\varepsilon)/\Delta$ and $x_4 = (-1-\varepsilon)/\Delta$. Now we make use of a quite general rule given in [14]: if the four roots $\alpha$, $\beta$, $\gamma$, $\delta$ of $X$ are all real, such that $\delta < \gamma < \beta < \alpha$, then the substitution

$$x = \frac{\gamma(\beta-\delta) - \delta(\beta-\gamma)\sin^2\varphi}{(\beta-\delta) - (\beta-\gamma)\sin^2\varphi}, \quad (6)$$

transforms

$$\frac{dx}{\sqrt{X}} \rightarrow \frac{2}{\sqrt{(\alpha-\gamma)(\beta-\delta)}} \frac{d\varphi}{\sqrt{1-m\sin^2\varphi}}, \quad (7)$$

where

$$m = \frac{(\beta-\gamma)(\alpha-\delta)}{(\alpha-\gamma)(\beta-\delta)} \text{ and } \gamma < x < \beta. \quad (8)$$

Now, for the integrals $g_I(\varepsilon>0; \Delta)$ and $g_{II}(\varepsilon>0; \Delta)$ we should attribute $x_1 \leftrightarrow \alpha$, $x_3 \leftrightarrow \beta$, $x_2 \leftrightarrow \gamma$, $x_4 \leftrightarrow \delta$ and $x_3 \leftrightarrow \alpha$, $x_1 \leftrightarrow \beta$, $x_2 \leftrightarrow \gamma$, $x_4 \leftrightarrow \delta$, respectively. Correspondingly, the both integrals finally read[#]

$$g_I(\varepsilon > 0;\Delta) = \frac{1}{2\gamma_1\pi^2} \sqrt{\Delta}\, K(m_I) \quad (9)$$

and

$$g_{II}(\varepsilon > 0;\Delta) = \frac{1}{2\gamma_1\pi^2} \frac{2\Delta}{\sqrt{(1+\Delta)^2 - \varepsilon^2}} K(m_{II}). \quad (10)$$

The $K(m_I)$ and $K(m_{II})$ are complete elliptic integrals of the first kind

$$K(m) = \int_0^{\pi/2} \frac{d\varphi}{\sqrt{1-m\sin^2\varphi}} \quad (11)$$

with the respective parameters

$$m_I = \frac{(1+\Delta)^2 - \varepsilon^2}{4\Delta} \text{ and } m_{II} = \frac{1}{m_I}. \quad (12)$$

It should be stressed that for $m_I$ the suitable range of $\varepsilon$ is $1-\Delta < \varepsilon < \Delta+1$ while for $m_{II}$ it equals to $0 < \varepsilon < 1-\Delta$, see Eq. (3). Thus, $0 < m_I < 1$ and also $0 < m_{II} < 1$, as expected. Of course, $m_I(\varepsilon>0; \Delta) = m_{IV}(\varepsilon<0; \Delta)$ and $m_{II}(\varepsilon>0; \Delta) = m_{III}(\varepsilon<0; \Delta)$.

In order to illustrate our results we need one thing more. Namely, as it was mentioned earlier, by the uniaxial pressure we expect some changes of the transfer integrals. Instead of using Harrison's relation [16] written in our case as $\Delta = (b/a)^{n_{12}}$ with the adjustable parameter $n_{12}$ we prefer

---

[#] The results were obtained in 1994 during stay at Gracefield Research Institute, Lower Hutt/Wellington.

consider both transfer integrals to be proportional to the overlap integral $S$ [17]

$$S(\rho) = \left(\frac{\rho^2}{3} + \rho + 1\right)\exp(-\rho) \qquad (13)$$

between two nearest-neighbours for atomic $s$-orbitals, i.e. $\gamma_1 \sim S(\rho_a)$ and $\gamma_2 \sim S(\rho_b)$. Here the proportionality factor can be set to 1 eV and the dimensionless nearest neighbour distances are defined in atomic units as $\rho_a \equiv a/a_0$ and $\rho_b \equiv b/a_0$, where $a_0 \cong 0.53$ Å means Bohr radius.

The overall density of states as a function of dimensionless energy $\varepsilon$ for various values of the lattice constant $b$ and fixed $a = 4$ at. u. $\cong 2.12$ Å is shown in Fig. 2.

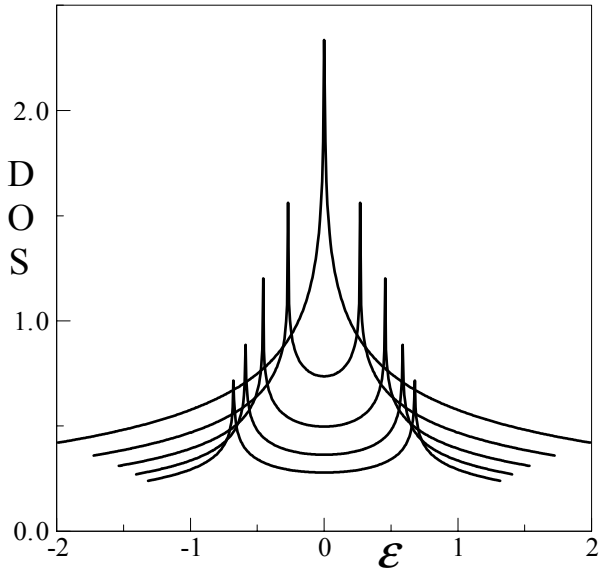

**Figure 2** Electron density of states in eV$^{-1}$ for various values of the coupled via Eq. (13) model parameters. From bottom to top, for a fixed $\gamma_1 = 1$ eV, we have $b = 2, 2.5, 3, 3.5$ and $3.998$ at. u. The values of the lattice deformation (or an anisotropy of the transfer integrals) measured by $1 - \Delta$ are correspondingly 0.677, 0.587, 0.457, 0.270 and 0.001. The logarithmic singularities appear at $\varepsilon = \pm(1 - \Delta)$.

The integrals $K(m_I)$ and $K(m_{II})$ were computed within polynomial approximation with the error not greater than $2.0 \times 10^{-8}$, see [15, 17.3.34].

## 3. Concluding remarks

The electron density of states given by Eqs. (9,10) for $\varepsilon \geq 0$ and by the corresponding symmetric parts for $\varepsilon < 0$ in a rectangular lattice behaves consistently with the limiting case of a square lattice, see the top curve in Fig. 2. On the other hand, for a fixed transfer integral $\gamma_1$ the larger $\gamma_2$ is, i.e. the larger lattice deformation measured by $1-\Delta$ is, the larger is separation distance equal to $2(1-\Delta)$ between the two singularities.

Finally, after this paper was completed the author became aware of Ref. [18], where a closed form of DOS in a rectangular lattice was also obtained but in a different context, not linked with the uniaxial pressure as we did.

**Acknowledgements**

I am indebted to Dr. Esther Haines for suggesting the question.